\documentclass[conference]{IEEEtran}
\usepackage{cite}

\usepackage{graphicx}
\ifCLASSINFOpdf
\else
\fi
\usepackage[cmex10]{amsmath}
\usepackage{array}

\usepackage{overpic}
\usepackage[T1]{fontenc}
\usepackage[utf8x]{inputenc}
\usepackage[english]{babel}
\usepackage{color}
\usepackage{soul}

	\newlength{\figurewidth}
	\newlength{\figurewidthA}
\begin{document}
%
\title{Inferring the interplay of network structure and market effects in Bitcoin}

\author{
\IEEEauthorblockN{D\'aniel Kondor, Istv\'an Csabai, J\'anos Sz\"ule, M\'arton P\'osfai and G\'abor Vattay}
\IEEEauthorblockA{Department of Physics of Complex Systems, E\"otv\"os Lor\'and University\\
H-1117, P\'azm\'any P\'eter s\'et\'any 1/A, Budapest, Hungary\\
Email: kdani88@elte.hu}
}


%


\maketitle

\begin{abstract}
	A main focus in economics research is understanding the time series of prices of goods and assets. While statistical
	models using only the properties of the time series itself have been successful in many aspects, we expect to gain a
	better understanding of the phenomena involved if we can model the underlying system of interacting agents.
	In this article, we consider the history of Bitcoin, a novel digital currency system, for which the complete list of
	transactions is available for analysis. Using this dataset, we reconstruct the transaction network between users
	and analyze changes in the structure of the subgraph induced by the most active users. Our approach is based on the
	unsupervised identification of important features of the time variation of the network. Applying the widely used
	method of Principal Component Analysis to the matrix constructed from snapshots of the network at different times,
	we are able to show how structural changes in the network accompany significant changes in the exchange price of
	bitcoins.
\end{abstract}



%
\IEEEpeerreviewmaketitle


\section{Introduction}

The growing availability of digital traces of human activity provide unprecedented opportunity for researchers to study complex
phenomena in society~\cite{happiness,szell,words,george,kallus,twitterpca}. Data mining methods that preform unsupervised
extraction of important features in large datasets are especially appealing because they enable researchers to identify patterns
without making a priori assumptions~\cite{twitterpca,lsalandauer}. In this article, we use Principal Component Analysis
(PCA)~\cite{pca} to study the dynamics of a network of monetary transactions. We aim to identify relevant changes in
network structure over time and to uncover the relation of network structure and macroeconomic indicators of the
system~\cite{changepoint,isvt,networkfinancial5,bubbles}.

In traditional financial systems, records of everyday monetary transactions are considered highly sensitive and are kept private.
Here, we use data extracted from Bitcoin, a decentralized digital cash system, where the complete list of such transactions is
publicly available~\cite{satoshi,shamir,anonimity}. Using this data, we construct daily snapshots of the transaction
network~\cite{kondor}, and we select the subgraph spanned by the most active users. We then represent the series of networks
by a matrix, where each row corresponds to a daily snapshot. We carry out PCA on this dataset identifying key features of
network evolution, and we link some of these variations to the exchange rate of bitcoins.

\section{Methods}

\subsection{The Bitcoin dataset}

	Bitcoin is a novel digital currency system that functions without central governing
	authority, instead payments are processed by a peer-to-peer network of users connected through the Internet. Bitcoin users announce new transactions
	on this network, and each node stores the list of all previous transactions. 
	
	Bitcoin accounts are referred to as \emph{addresses} and consist of a pair of public and private keys. One can receive bitcoins by providing the public
	key to other users, and sending bitcoins requires signing the announced transaction with the private
	key. Using appropriate software\footnote{several open-source Bitcoin clients exist, see
	e.g.~\texttt{https://bitcoin.org/en/choose-your-wallet}}, a user can generate unlimited number
	of such addresses; using multiple addresses is usually considered a good practice to increase a user's privacy. To support this,
	Bitcoin transactions can have multiple \emph{input} and \emph{output} addresses; bitcoins provided by the input addresses can
	be distributed among the outputs arbitrarily.
	
	Due to the nature of the system, the record of all previous transactions since its beginning are publicly
	available to anyone participating in the Bitcoin network. We installed a slightly modified version
	of the open-source \texttt{bitcoind} client and downloaded the list of transactions from the peer-to-peer network on March 3th, 2014.
	From these records, we recovered the sending and receiving addresses, the sum involved and the approximate time of
	each transaction.   Throughout the paper, we only analyze transactions which happened in 2012 or 2013.
	We made the dataset and the source code of the modified client program available
	on the project's website~\cite{web} and through our online database interface~\cite{nmvo,matray}.
	
\subsection{Extracting the core network}

	To study structural changes in the network and their connection with outside events, we extract the subgraph of the most active
	Bitcoin users. As a first step, we apply a simple heuristic approach to identify addresses belonging to the same user~\cite{anonimity}.
	We call this step the \emph{contraction} of the graph. For this, we identify all transactions with multiple input addresses
	and consider addresses appearing in the same transaction to belong to the same user. Since initiating a
	transaction requires signing it with the private keys of all input addresses, we expect that these are controlled by the same entity. Note that this procedure will fail to contract addresses that belong to the same user, but are used completely independently. However, this is the most widely accepted method~\cite{anonimity,shamir,bitiodine}. Therefore each node $v$ in the transaction network represents a user, and each link $(u\to v)$ represents that there was at least one transaction between users $u$ and $v$ during the observed two-year period.
	
	We identify the active core of the transaction graph using two different approaches:
	 (i)~We include users who appear in at least 100 individual
	transactions and were active for at least 600 days, i.e.~at least 600 days passed between their first and last
	appearance in the dataset. We call these \emph{long-term} users, and we refer to the extracted network as the LT core.
	 (ii)~We simply include the 2,000 most active
	users. All users are considered, hence the resulting network is referred to as AU core. In both cases, we take the largest connected component of the graph induced by the selected nodes.
	Furthermore, we exclude nodes which are associated with the SatoshiDice gambling
	site\footnote{\texttt{http://satoshidice.com}; their addresses used for the service start with `\texttt{1Dice}'}.
	In 2012, this site and its users produced over half of all Bitcoin activity, 
	which is not related to the normal functioning of the system.
	
	In the case of long-term users, the LT core consists of $n_\text{LT}=1,639$ nodes and $l_\text{LT}=4,333$ edges;
	these users control a total of 1,894,906 Bitcoin addresses which participated in 4,837,957 individual transactions
	during the examined two-year period. In the case of most active users, the AU core has $n_\text{AU}=1,288$ nodes
	and $l_\text{AU}=7,255$ edges; the users in this subgraph have a total of 3,326,526 individual Bitcoin addresses
	and participated in a total of 12,900,964 transactions during the two years. The total number of Bitcoin transactions
	in this period is 27,930,580, meaning that the two subgraphs include $17.3\%$ and $46.2\%$ of all transactions
	respectively.

\subsection{Detecting structural changes}

To extract important changes in the graph structure, we compare successive snapshots of the active core of the transaction network using PCA. The goal is to obtain a set of \emph{base networks}, and represent each day's snapshot as a linear combination of these base networks.

We calculate the daily snapshots of the active core, each snapshot is a weighted network, and the weight of link $(u\rightarrow v)$ is equal to the number of transactions that occurred that day between $u$ and $v$. A snapshot for day $t$ can be represented by an $n\times n$ weighted adjacency matrix $W_t$, where $n \equiv n_\text{LT/AU}$ is the number of nodes in the aggregate network.
Since there are $l \equiv l_\text{LT/AU}$ links overall, each $W_t$ has at maximum $l$ nonzero elements. We rearrange $W_t$ into an $l$ long vector $w_t$. Note that we include all possible links, even if for that specific day, some are missing. For $T$ snapshots, we now construct the $T \times l$ \emph{graph time series matrix} $X$ such that the $t$th row of $X$ equals $w_t$~\cite{isvt}. This way, we can consider $X$ as a data matrix with $T$ samples and $l$ features, i.e.~we consider each day as a sample, and the activities of possible edges as features.

To account for the for the high variation of the daily number of transactions, we normalize $X$ such that the sum of each row equals $1$. 
After that, as usual in PCA, we subtract the column averages from each column.
As a result, both the row and column sums in the matrix will be zero. We compute the singular value decomposition (SVD)
of the matrix $X$:
\begin{equation}
	X = U \Sigma V^{T},
	\label{eqsvd}
\end{equation}
where $\Sigma$ is a $T \times l$ diagonal matrix containing the singular values (which are all positive),
and the columns of the $T \times T$ matrix $U$ and the $l \times l$ matrix $V$ contain the singular vectors.
Since in our case $T < l$, there will be only $T$ nonzero singular values and only $T$ relevant singular vectors;
as usual, we keep only the relevant parts of the matrices, this way $\Sigma$ will be only $T \times T$ and $V$ will be
only $l \times T$~\cite{numrec}. The singular vectors are orthonormal, i.e.~$U^{T} U = V V^{T} = I$, where $I$ is the $T \times T$
identity matrix. It is customary to order the singular values and vectors such that the singular values are in decreasing
order, so that the successive terms in the matrix multiplication in \eqref{eqsvd} give decreasing contribution to the
result, thus giving a successive approximation of the original matrix. Note that these matrices can also be computed
as the eigenvectors of the covariance matrices: $X X^T$ and $X^T X$ for $U$ and $V$ respectively, and as such, and the
columns of $U$ and $V$ span the row and column space of $X$. In accordance with this, we can interpret the singular vectors
based on the interpretation of $X$.
The columns of $V$ can be considered as `base networks', the matrix element $v_{ji}$ provides the weight of edge $j$ in base network $i$.
According to \eqref{eqsvd}, edge weights in the daily snapshots can be calculated as a linear combination of edge weights in the
base networks. The singular values give the overall importance of a base network in approximating the whole time series,
while the columns of $U$ account for the temporal variation: the matrix element $u_{ti}\equiv u_i(t)$ provides the contribution
of base network $i$ at time $t$, e.g.~the (normalized) weight of edge $j$ on day $t$ is given by:
\begin{equation}
	x_{tj} = \sum_{i=1}^T \Sigma_{ii} u_j (t) v_{ji}.
	\label{eqxtj}
\end{equation}

\section{Results}

\begin{figure}
	\centering
	\begin{overpic}[width=\figurewidth 
		]{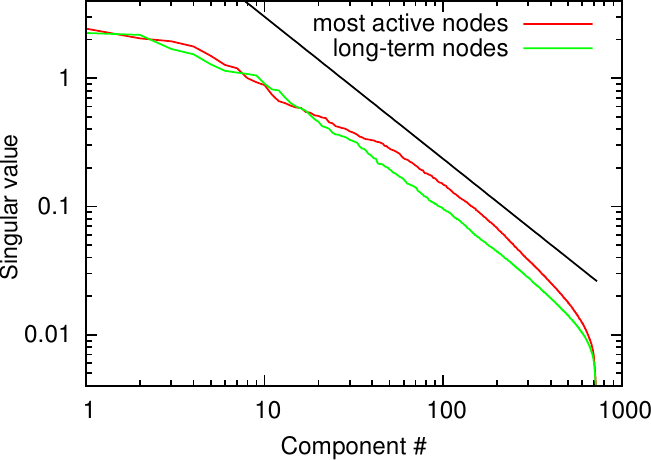}
			\put(8,11){\includegraphics[width=4cm]{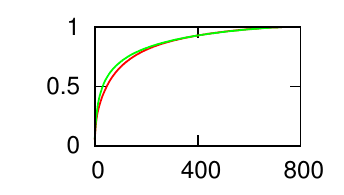}}
	\end{overpic}
	\caption{Singular values of $X$. The inset shows the relative contribution up to the given index.
		The distribution of singular values is fat-tailed, a high number of components is required to explain the variations in the
		data. The best power-law fit is $\Sigma_{ii} \sim i^{-1.37}$.}
	\label{ns1s}
\end{figure}

\begin{figure*}
	\centering
	\includegraphics[width=\figurewidth]{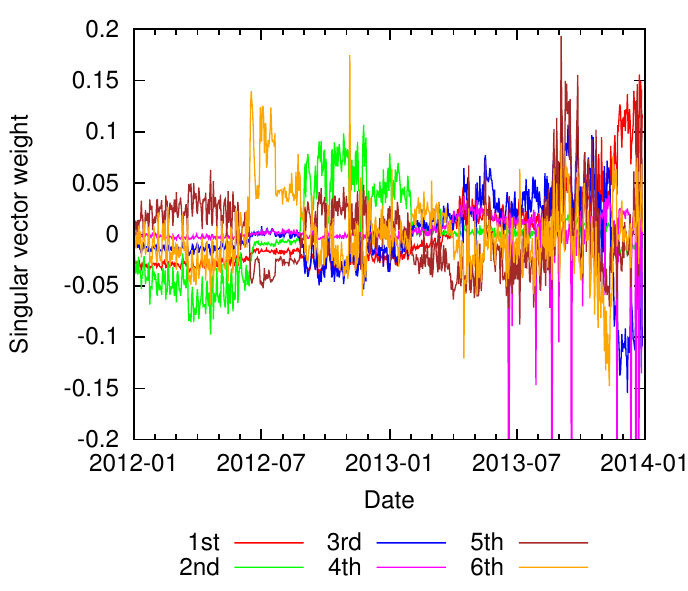}
	\quad
	\includegraphics[width=\figurewidth]{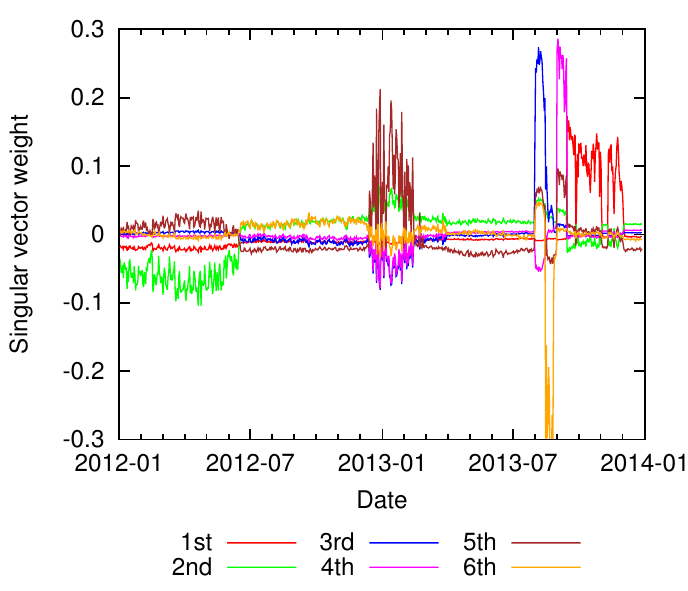}
	
	\caption{The time-varying contribution $u_i(t)$ of the first $6$ base networks, for the LT core (left) and AU core (right).}
	\label{nu14}
\end{figure*}

\begin{figure*}
	\centering
	\begin{overpic}[width=\figurewidthA 
		]{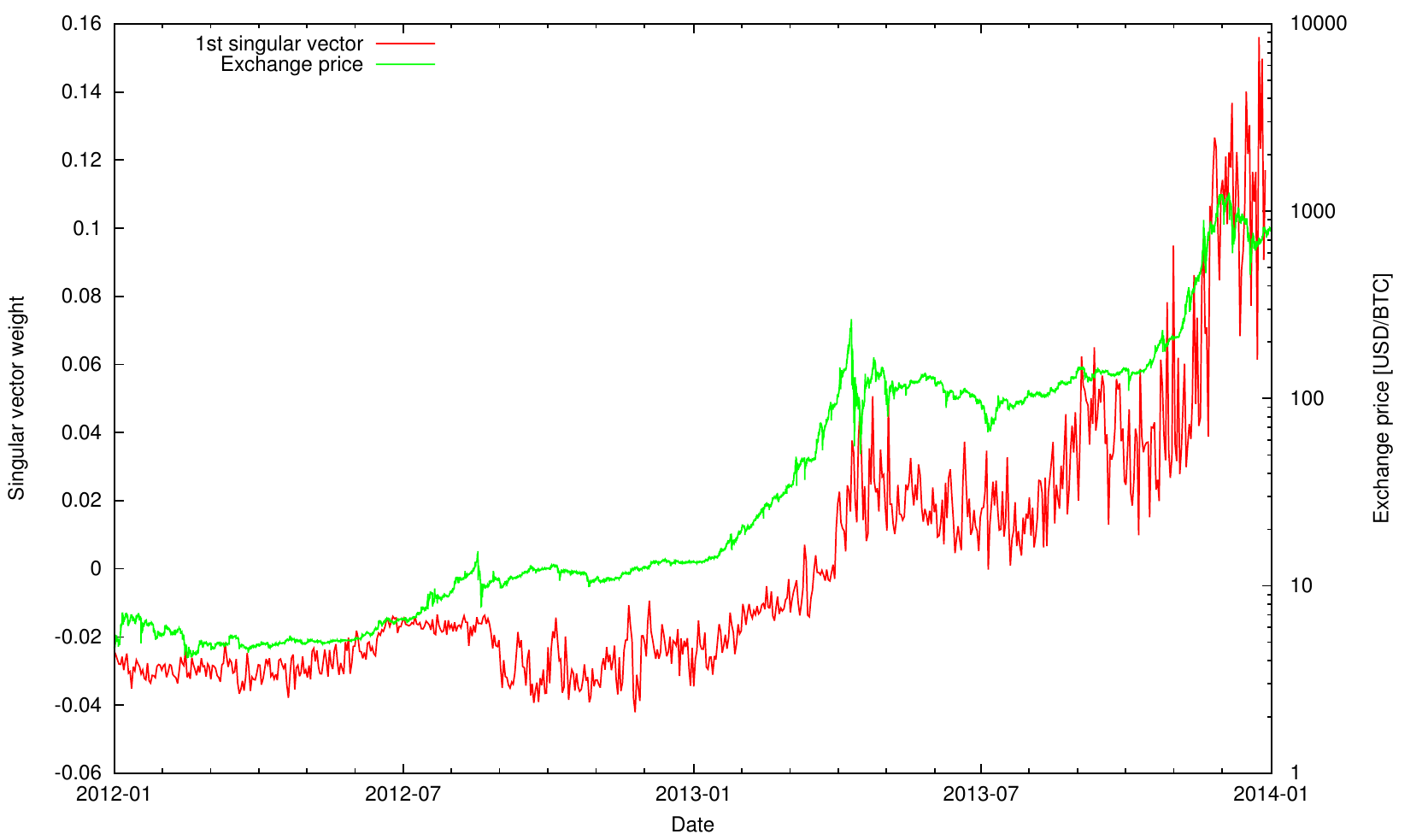}
			\put(8,20){\includegraphics [width=3cm]{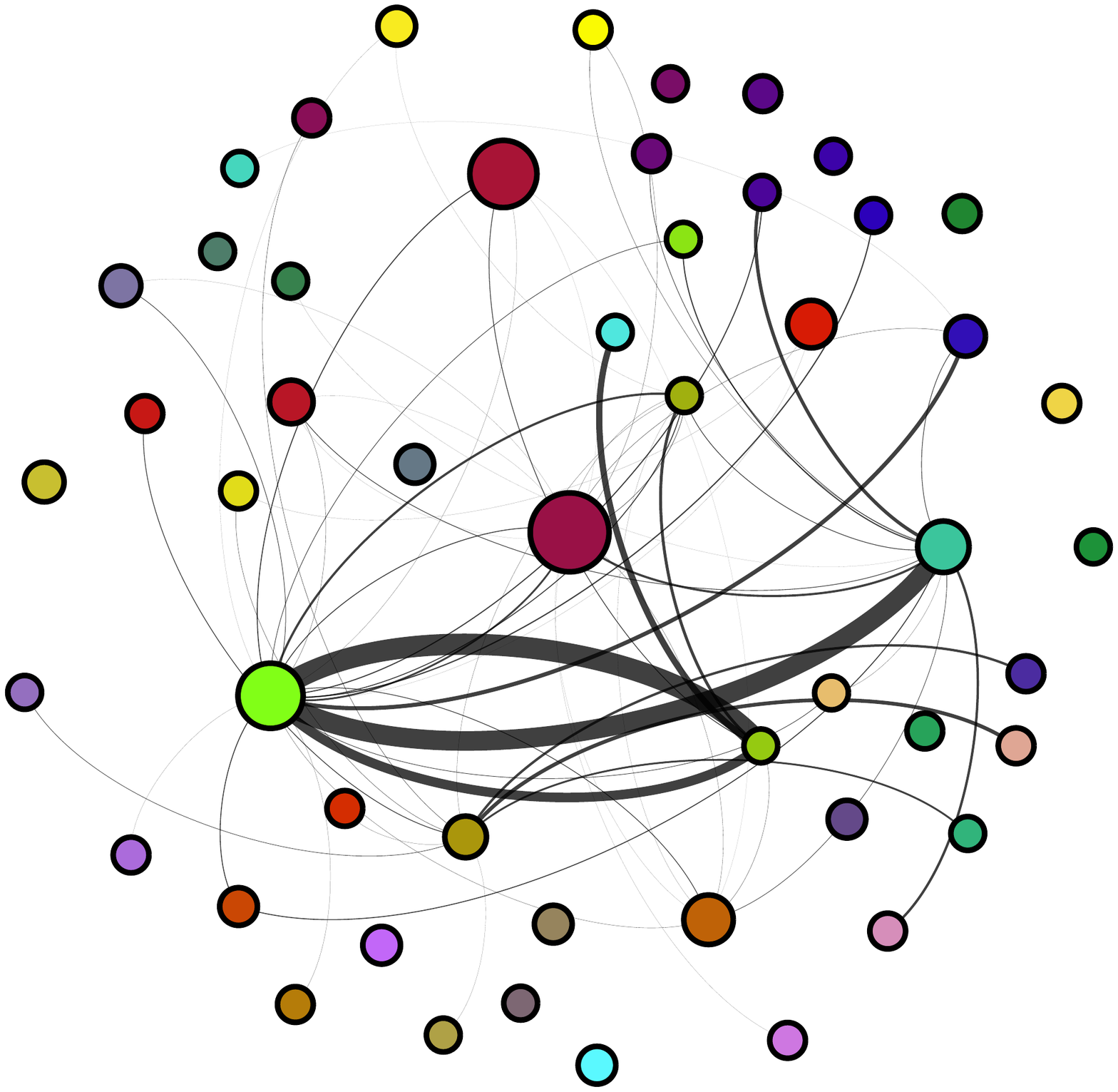}}
			\put(25,25){\includegraphics[width=3cm]{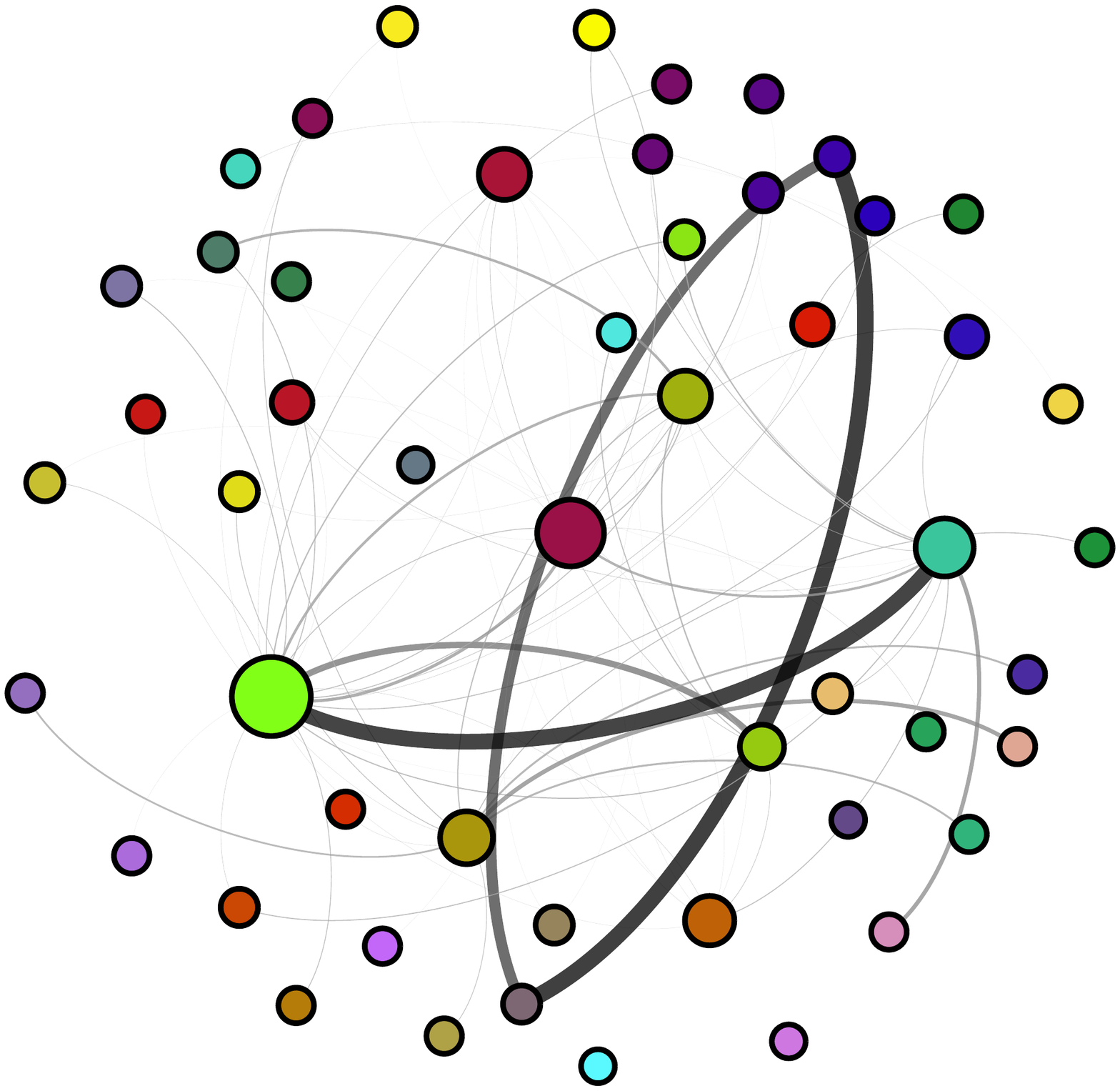}}
			\put(60,30){\includegraphics[width=3cm]{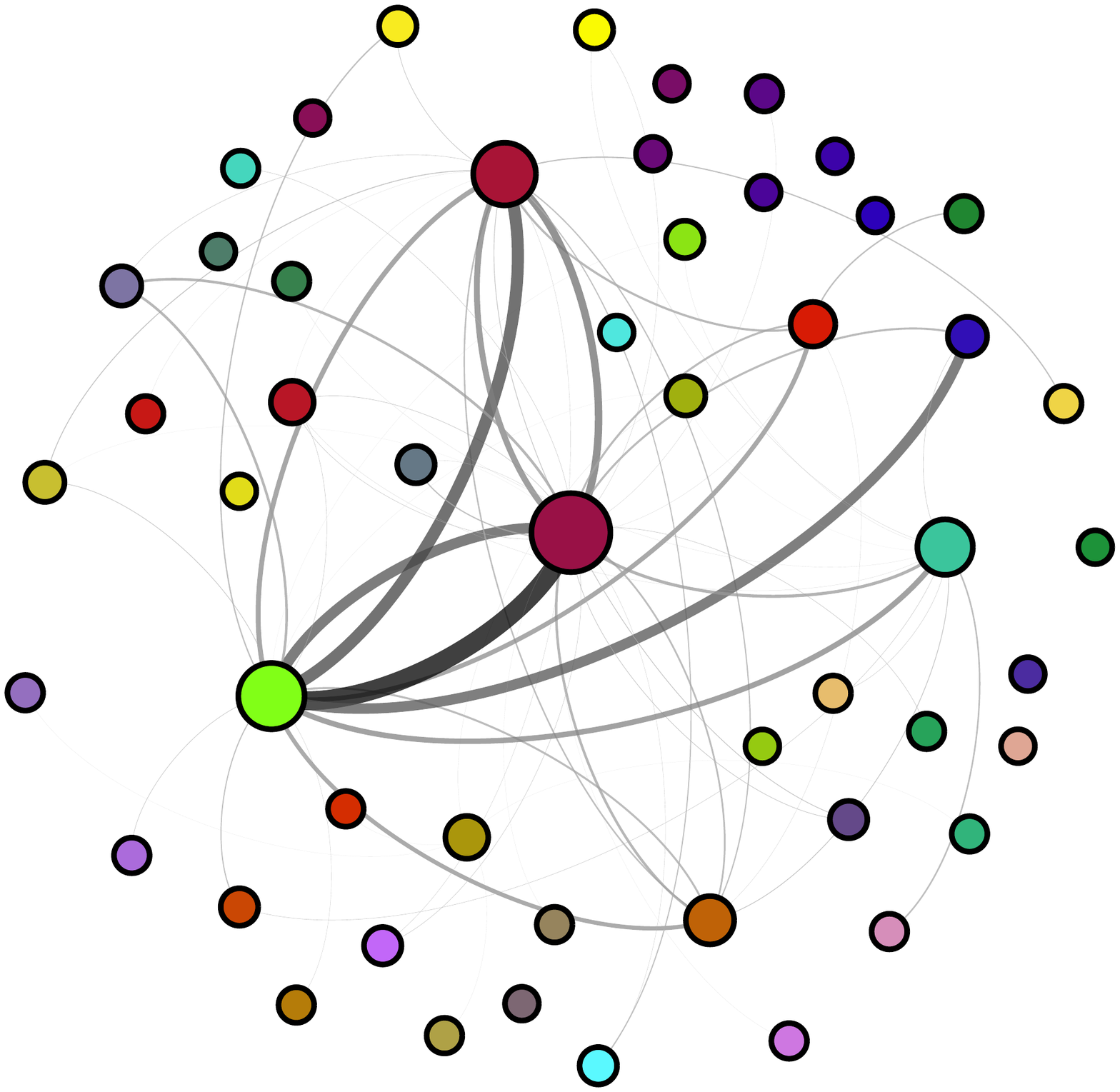}}
			\put(72,1){\includegraphics [width=3cm]{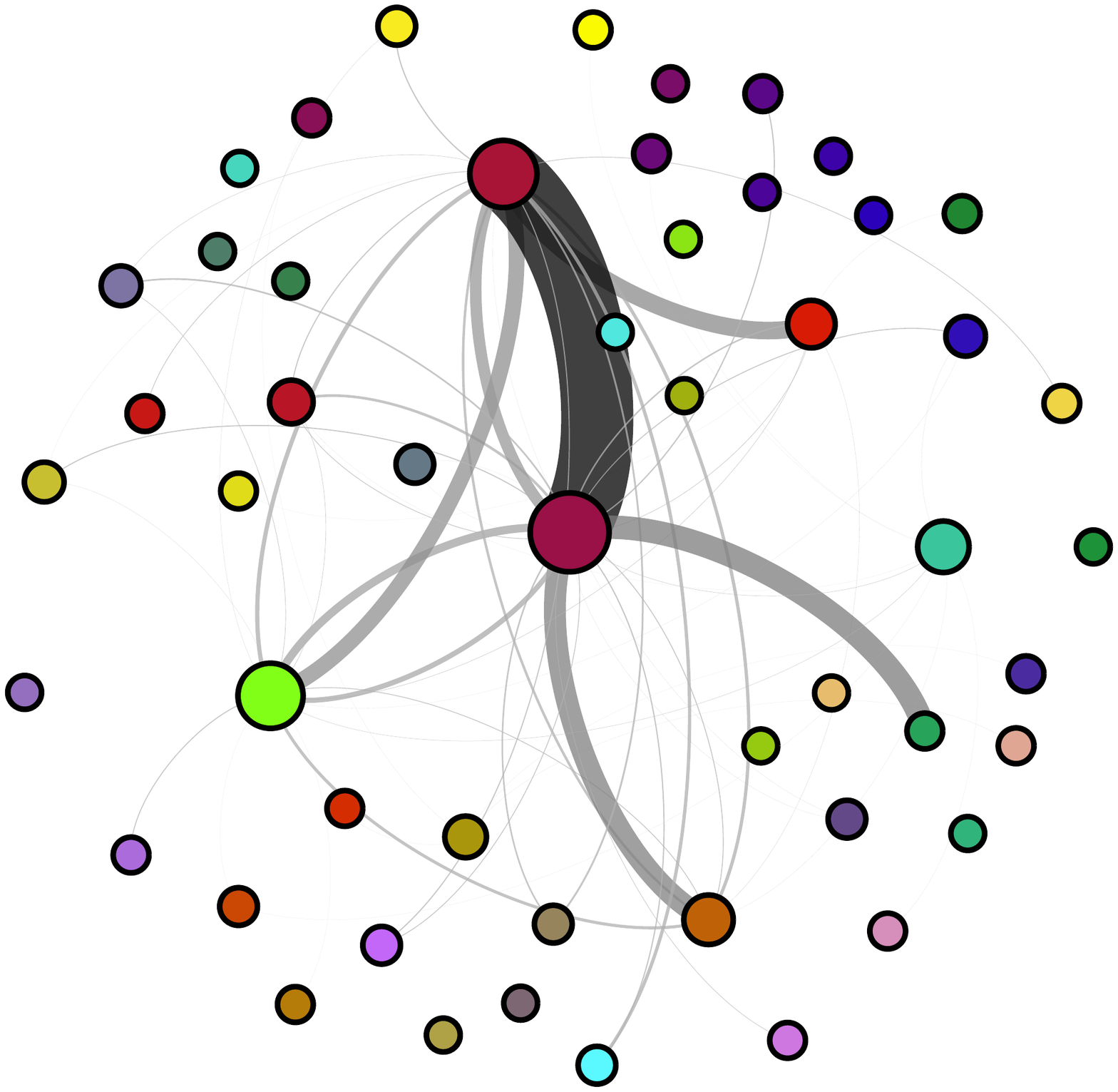}}
	\end{overpic}
	\caption{The time-varying contribution of the first base network $u_1(t)$ for the LT core and the time series of the logarithm
		of exchange price at the MtGox trading site. To illustrate how this corresponds to changes in the network, snapshots of
		the subnetworks induced by top 100 edges with the largest absolute weight in the base network are also shown for a few
		dates. Visual inspection shows a good correspondence of the two; also, the two datasets have high correlation coefficients,
		see Table~\ref{tcorr}.}
	\label{u2}
\end{figure*}

Examining the singular values, we find that for both type of core extraction they decay only relatively slowly,
i.e.~a large number of components are required to obtain a good approximation of the original matrix $X$ (see Fig.~\ref{ns1s}).
This indicates that the system possibly contains non-Gaussian noise and high dimensional structure.
Also, the distribution of edge weights $p_i(v)$ in the base networks is fat-tailed; for the first
six base networks we find very similar distributions, all well approximated with $p_i(v) \sim v^{-1.8}$
for the LT core and $p_i(v) \sim v^{-1.9}$ for the AU core.

Examining the edges with large weights for the LT core, we find that most of these
are repeated within the first few base networks. For example, if we consider the top-20 ranking edges (by the absolute value of weights)
in the first 10 base networks, we find only 46 distinct edges instead of the 200 maximally possible.
Among these, 44 induce a weakly connected graph, including a total of 29 users; considering
all edges among these users, 20 of them form a strongly connected component by themselves
and all are weakly connected. These 29 users have a total of 1,349,815 separate Bitcoin addresses,
forming a highly active subset in our core network.

We show the time-varying contribution $u_i(t)$ of the first six base networks on Fig.~\ref{nu14}. 
In most cases, $u_i(t)$ features a few abrupt changes, partitioning the history of Bitcoin
into separate time periods. This is especially true for the AU core, where highly active but short lived users
can significantly contribute. Identifying the individual nodes and edges responsible for network activity in a given period would
require more information about Bitcoin addresses, which is difficult to obtain on a large scale.

The most striking feature uncovered by our analysis is a clear correspondence between the first singular vector of the graph of long-term
users and the market price of bitcoins as shown on Fig.~\ref{u2}. Apart from visual similarity, the two datasets have a
significantly high correlation coefficients (see Table~\ref{tcorr}).

\begin{figure}
	\centering
	\includegraphics[width=\figurewidth]{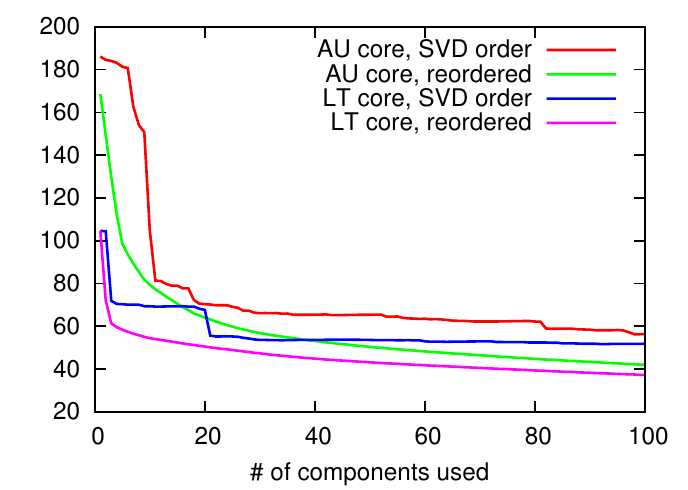}
	
	\caption{Residual standard error of approximating the exchange rate of bitcoins with a linear combination of the time-varying contribution of base networks.}
	\label{lmerr}
\end{figure}

\begin{figure*}
	\centering
	\includegraphics[width=\figurewidth]{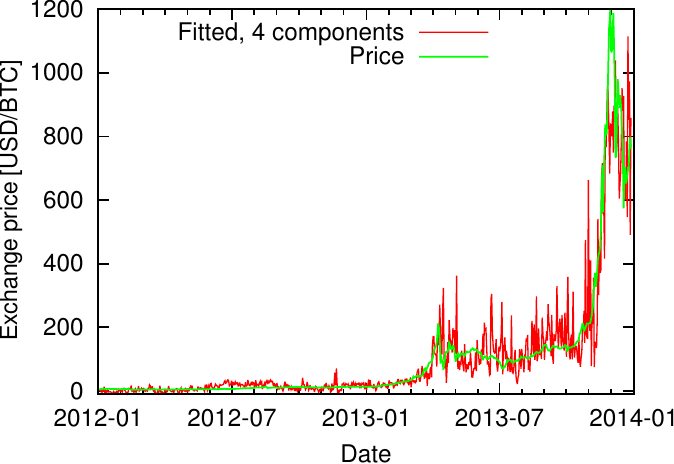}
	\quad
	\includegraphics[width=\figurewidth]{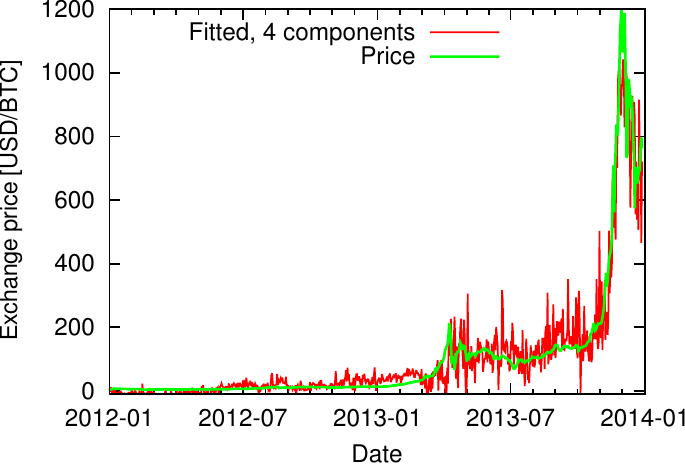}
	
	\caption{Approximating the exchange price of bitcoins with a linear combination of singular vectors of the time series matrix
		of the graph induced by the long-term nodes. Left: using the first $4$ vectors ranked by singular values;
		right: using the first $4$ vectors ranked by Pearson correlation with the price time series.}
	\label{lmfit}
\end{figure*}

Motivated by this result, we tested whether the price of bitcoins can be estimated with the $u_i(t)$ coefficients. To proceed, we
subtract the average value from the price time series, and estimate this as a linear combination of singular vectors:
\begin{equation}
	\tilde{B}(t) = \left\langle B(t)\right\rangle + \sum_{i=1}^N c_i u_i(t) \, \textrm{.}
\end{equation}
Here the coefficients $c_i$ can be computed as the dot product of the price time series and $u_i(t)$, and therefore $c_i$ is proportional 
to the Pearson correlation coefficients shown in Table~\ref{tcorr}. We are interested how large $N$ is needed to model the price time
series with acceptable accuracy. We show the residual standard error as a function of $N$ on
Fig.~\ref{lmerr}. It is apparent that there are sudden decreases in the error after the inclusion of certain
base networks. We note that the base networks are ranked by their contribution to the original network time series matrix,
i.e.~the singular values of $X$. On the other hand, we can rank these components by their similarity to the price time
series by using correlation coefficients. This results in a more rapid decrease of
the error, and the large jumps are all at the beginning. In Table~\ref{tcorr}, we show the first few components with the
highest correlation coefficients. Two approximations of the time series for
the LT core is shown on Fig.~\ref{lmfit}; one with the first $4$ base networks, the other with the $4$
base networks whose time-varying contribution $u_i(t)$ has the highest correlation to the price. We find that in both cases most features of the time series are
well approximated, but the fitted time series is still apparently noisy. In the first case, the shape of the peak at the end
of 2013 is missed, while in the second case it is approximated much better. A closer examination of the correlation values
(Table~\ref{tcorr}) and the singular vectors reveals that the 21st component is responsible for this change, which contains
high resemblance to the final section of the time series.

\begin{table*}
\centering
\begin{tabular}{r|r|r|r|r}
		& \multicolumn{2}{c|}{long-term nodes} & \multicolumn{2}{c}{most active nodes} \\
	Component & $\rho_P$ & $\rho_S$ & $\rho_P$ & $\rho_S$ \\ \hline
	$1$  &  $0.8528$	&  $0.8654$ &  $0.3689$ &  $0.9231$ \\
	$2$  &  $-0.0335$	&  $0.4367$ &  $0.1233$ &  $0.4435$ \\
	$3$  &  $-0.3782$	&  $0.4407$ &  $0.0779$ & $-0.0268$ \\
	$4$  &  $-0.074$	&  $0.5913$ &  $0.0972$ &  $0.3568$ \\
	$5$  &  $-0.0148$	& $-0.2991$ & $-0.1313$ & $-0.3066$ 
\end{tabular}
\,
\begin{tabular}{r|r|r|r|r|r}
	\multicolumn{3}{c|}{long-term nodes} & \multicolumn{3}{c}{most active nodes} \\
	Component & $\rho_P$ & $\rho_S$ & Component &$\rho_P$ & $\rho_S$ \\ \hline
	$1$  & $0.8528$  & $0.8654$ & $10$ & $-0.5395$ & $0.0075$  \\
	$3$  & $-0.3782$ & $0.4407$ & $7$  &  $0.3928$ & $0.3633$  \\
	$21$ & $0.1898$  & $0.0466$ & $1$  &  $0.3689$ & $0.9231$  \\
	$4$  & $-0.074$  & $0.5913$ & $11$ &  $0.3276$ & $-0.0578$ \\
	$19$ & $0.0589$  & $0.0423$ & $8$  & $-0.2632$ & $-0.3631$ 
\end{tabular}
\caption{Correlation coefficients between the singular vectors of the network time series matrix and the Bitcoin exchange price.
	Left: first $5$ singular vectors. Right: the $5$ singular vectors with the highest correlation. Here, $\rho_P$ is the
	Pearson correlation coefficient and $\rho_S$ is the Spearman rank correlation coefficient.}
\label{tcorr}
\end{table*}

\section{Conclusions and future work}

In this paper, we investigated whether connection between the network structure and macroscopic properties (i.e.~the exchange price)
can be established in the Bitcoin network. For our analysis, we reconstructed daily network snapshots of the networks of the most
active users in a two-year period. We organized these snapshots into the graph time series matrix of the system. We analyzed this matrix
using PCA which allowed us to identify changes in the network
structure. A striking feature we found was that the time-varying contribution of some of the base networks show a clear
correspondence with the market price of bitcoins. The contribution of the first base network was found to be exceptionally similar. Using the linear
combination of only $4$ vectors, we were able to reproduce most of the features of the long-term time evolution of the market price.

Based on our results, it is apparent that the analysis of the structure of the underlying network of a financial system can provide
important new insights complementing analysis of the external features. Further research could focus on establishing causal relationship between the observed features, and possibly predicting price changes based
on structural changes in the network. Also, collecting publicly available information about Bitcoin addresses identified
as members of the highly active core of the network could result in a better understanding of the role of the associated users in the Bitcoin ecosystem, and help explain the correlations observed here.

\section*{Acknowledgments}

	The authors thank the partial support of
	the European Union and the European Social Fund through project FuturICT.hu
	(grant no.: TAMOP-4.2.2.C-11/1/KONV-2012-0013), the OTKA 7779 and the NAP 2005/KCKHA005 grants.
	EITKIC\_12-1-2012-0001 project was partially supported by the Hungarian Government,
	managed by the National Development Agency, and financed by the Research and
	Technology Innovation Fund and the MAKOG Foundation.


\end{document}